\documentclass[aps,prd,nofootinbib,showpacs,superscriptaddress,preprint]{revtex4-1}
\pdfoutput=1
\usepackage[utf8]{inputenc}
\usepackage{amsmath,amssymb}
\usepackage{graphicx}
\usepackage{multirow}
\usepackage[usenames,dvipsnames]{color}
\usepackage{float}
\usepackage[colorlinks,citecolor=blue]{hyperref}
\usepackage[compat=1.0.0]{tikz-feynman}
\usepackage{color}
\usepackage{subfigure}
\DeclareUnicodeCharacter{2212}{-}

\begin{document}
\title{Anomalous Magnetic Moment and Higgs Coupling
of the Muon in a Sequential U(1) Gauge Model with Dark Matter}

\author{Rathin Adhikari}
\email{rathin@ctp-jamia.res.in}
\affiliation{Centre for Theoretical Physics, Jamia Millia Islamia - Central University, Jamia Nagar, New Delhi - 110025, India}

\author{Imtiyaz Ahmad Bhat}
\email{imtiyaz@ctp-jamia.res.in}
\affiliation{Centre for Theoretical Physics, Jamia Millia Islamia - Central University, Jamia Nagar, New Delhi - 110025, India}

\author{Debasish Borah}
\email{dborah@iitg.ac.in}
\affiliation{Department of Physics, Indian Institute of Technology Guwahati, Assam 781039, India}

\author{Ernest Ma}
\email{ma@physastro.ucr.edu}
\affiliation{Department of Physics and Astronomy,
University of California, Riverside, California 92521, USA}

\author{Dibyendu Nanda}
\email{psdn2502@iacs.res.in}
\affiliation{School of Physical Sciences, Indian Association for the Cultivation of Science,2A \& 2B Raja S. C. Mullick Road, Kolkata 700032, India}

\begin{abstract}
We study an Abelian gauge extension of the standard model with fermion families having non-universal gauge charges. The gauge charges and scalar content are chosen in such an anomaly-free way that only the third generation fermions receive Dirac masses via renormalisable couplings with the Higgs boson. Incorporating additional vector like fermions and scalars with appropriate $U(1)$ charges can lead to radiative Dirac masses of first two generations with neutral fermions going in the loop being dark matter candidates. Focusing on radiative muon mass, we constrain the model from the requirement of satisfying muon mass, recently measured muon anomalous magnetic moment by the E989 experiment at Fermilab along with other experimental bounds including the large hadron collider (LHC) limits. The anomalous Higgs coupling to muon is constrained from the LHC measurements of Higgs to dimuon decay. The singlet fermion dark matter phenomenology is discussed showing the importance of both annihilation and coannihilation effects. Incorporating all bounds lead to a constrained parameter space which can be probed at different experiments.
\end{abstract}

\maketitle

\section{Introduction}
\label{sec:intro}
Recently, the E989 experiment at Fermilab has measured the anomalous magnetic moment (AMM) of muon, $a_\mu$ = $(g - 2)_\mu/2$, showing a  discrepancy with respect to the theoretical prediction of the Standard
	Model (SM) \cite{Abi:2021gix}
	\begin{eqnarray}
		a^{\rm FNAL}_\mu = 116 592 040(54) \times 10^{-11}\\
		a^{\rm SM}_\mu = 116 591 810(43) \times 10^{-11}.
	\end{eqnarray}
When combined with the previous Brookhaven determination
	\begin{equation}
		a^{\rm BNL}_\mu = 116 592 089(63) \times 10^{-11},
	\end{equation}
	it leads to a 4.2 $\sigma$ observed excess of
	$\Delta a_\mu = 251(59) \times 10^{-11}$ \footnote{It should however, be noted that the latest lattice results \cite{Borsanyi:2020mff} predict a larger value of muon $(g-2)$ bringing it closer to experimental value. Tension of measured muon $(g-2)$ with global electroweak fits from $e^+ e^-$ to hadron data was also reported in \cite{Crivellin:2020zul, Colangelo:2020lcg, Keshavarzi:2020bfy}.}. The theoretical status of SM calculation of muon AMM can be found in  \cite{Aoyama:2020ynm}. While this anomaly is known for a long time since the Brookhaven measurements \cite{Muong-2:2001kxu}, the recent Fermilab measurements  have also led to several recent works on updating possible theoretical models with new data, a comprehensive review of which may be found in \cite{Athron:2021iuf}. Earlier reviews on this topic can be found in \cite{Jegerlehner:2009ry, Lindner:2016bgg}.

	In this work, we consider an anomaly free $U(1)_X$ gauge extension of the SM where first two generations of charged fermions acquire masses only at radiative level. While triangle anomalies cancel due to addition of chiral fermion triplets, giving rise to type III seesaw origin of light neutrino masses, the new fields introduced for radiative charged fermion masses can also serve as a stable dark matter (DM) candidate, if it is stable and neutral. Focusing primarily on radiative muon mass and muon AMM, we constrain the model from the requirement of satisfying muon mass, latest muon $(g-2)$ data along with other relevant bounds like the Higgs coupling to muons as measured by the large hadron collider (LHC), Higgs to diphoton bound as well as direct search bounds on beyond standard model (BSM) particles. We also constrain the model from the requirement of generating the desired DM phenomenology. Radiative charged lepton mass in the context of AMM have been a topic of interest for many years and several interesting works have already appeared in the literature within supersymmetric \cite{Borzumati:1999sp, Czarnecki:2001pv, Crivellin:2010ty, Thalapillil:2014kya} as well as non-supersymmetric frameworks \cite{Fraser:2014ija, Fraser:2015zed, Calibbi:2020emz, Yin:2021yqy, Chiang:2021pma, Baker:2021yli}. On the other hand, connection between dark matter and muon $(g-2)$ have also been studied in several earlier works, but with tree level muon mass \cite{Calibbi:2018rzv, Kawamura:2020qxo, Chen:2020tfr, Jana:2020joi, Kowalska:2017iqv, Kowalska:2020zve, Arcadi:2021cwg, Chowdhury:2021tnm}.

	We provide a natural origin of muon AMM together with radiative muon mass and dark matter in a sequential $U(1)_X$ gauged model that can also explain light neutrino mass from type III seesaw. The particle content and the corresponding $U(1)_X$ charge assignments are chosen in such an anomaly free way that additional global symmetries are not required. The radiative muon mass leads to anomalous Higgs coupling to muon which can be probed at the LHC. In spite of having several BSM particles and free parameters, we find the model to be highly constrained from the requirements of satisfying relevant constraints. 
	
	This paper is organised as follows. In section \ref{sec:model}, we briefly discuss the model. In section \ref{sec:g-2}, we discuss the possible origin of muon $(g-2)$ in this model followed by discussion of electroweak precision constraints in section \ref{sec:S&T}. We briefly comment upon electric dipole moment and lepton flavour violation constraints in section \ref{sec:edmlfv} followed by discussion of collider constraints in section \ref{sec:lhc}. In section \ref{sec:DM} we discuss DM details and summarise our results in section \ref{sec:conclude}.

\section{The Model}
\label{sec:model}

\begin{center}
\begin{table}\label{table1}
\caption{Fermion Content of the minimal model}
\begin{tabular}{|c|c|c|}
\hline
Particle & $SU(3)_c \times SU(2)_L \times U(1)_Y$ & $U(1)_X$  \\
\hline
$ (u,d)_L $ & $(3,2,\frac{1}{6})$ & $n_1$ \\
$ u_R $ & $(\bar{3},1,\frac{2}{3})$ & $\frac{1}{4}(7 n_1 -3 n_4)$  \\
$ d_R $ & $(\bar{3},1,-\frac{1}{3})$ & $\frac{1}{4} (n_1 +3 n_4)$ \\
$ (\nu, e)_L $ & $(1,2,-\frac{1}{2})$ & $n_4$  \\
$e_R$ & $(1,1,-1)$ & $\frac{1}{4} (-9 n_1 +5 n_4)$ \\
\hline
$\Sigma_{R} $ & $(1,3,0)$ & $\frac{1}{4}(3n_1+n_4)$ \\
\hline
\end{tabular}
\end{table}
\end{center}
The fermion content of the minimal model is shown in table \ref{table1}. The $U(1)_X$ charges correspond to anomaly-free combination with $n_1, n_4$ being arbitrary with $n_4 \neq -3n_1$. While such Abelian extension of the standard model was studied before \cite{Ma:2002pf, Barr:2005je, Adhikari:2008uc, Adhikari:2015woo, Bhat:2019yqo} in different contexts, recently the possibility of having a sequential $U(1)_X$ with different quantum numbers for each family was proposed \cite{Ma:2021aag}. As an working example, $n_1=0$ for all three families while $n_4=2, 1, 0$ for first, second and third families respectively were chosen. Now, if just one scalar doublet is chosen having zero $U(1)_X$ charge and responsible for electroweak symmetry breaking, only the third generation quarks and charged leptons can acquire masses at renormalisable level\footnote{See \cite{Balakrishna:1987qd, Balakrishna:1988ks, Ma:1990ce, Barr:1989ta, Babu:1988fn, Weinberg:2020zba} for earlier discussions on fermion mass hierarchy through sequential loop suppression.}. The field content of the minimal model with such choices of $n_1, n_4$ is shown in table \ref{table2}. As discussed in \cite{Ma:2021aag}, such a minimal setup leads to tree level third generation charged fermion masses while the first and second generation masses arise only at dimension six and dimension five levels, leading to natural suppression. 

\begin{center}
\begin{table}\label{table2}
\begin{tabular}{|c|c|c|}
\hline
Particle & $SU(3)_c \times SU(2)_L \times U(1)_Y$ & $U(1)_X$  \\
\hline
$ (u,d)_L, (c,s)_L, (t,b)_L $ & $(3,2,\frac{1}{6})$ & $0$ \\
$ u_R, c_R, t_R $ & $(\bar{3},1,\frac{2}{3})$ & $-\frac{3}{2}, -\frac{3}{4}, 0$  \\
$ d_R, s_R, b_R $ & $(\bar{3},1,-\frac{1}{3})$ & $\frac{3}{2}, \frac{3}{4}, 0$ \\
$ (\nu_e, e)_L, (\nu_{\mu}, \mu)_L, (\nu_{\tau}, \tau)_L $ &  $(1,2,-\frac{1}{2})$ & $2, 1, 0$  \\
$e_R, \mu_R, \tau_R$ & $(1,1,-1)$ & $\frac{5}{2}, \frac{5}{4}, 0$ \\
\hline
$\Sigma^e_{R}, \Sigma^{\mu}_{R}, \Sigma^{\tau}_{R} $ & $(1,3,0)$ & $\frac{1}{2}, \frac{1}{4}, 0$ \\
\hline
$\Phi=(\phi^+, \phi^0)$ & $(1,2,\frac{1}{2})$ & $0$ \\
$\eta_1, \eta_2$ & $(1,1,0)$ & $\frac{1}{4}, \frac{3}{4}$ \\
\hline
\end{tabular}
\caption{Particle content of the minimal model with chosen $n_1, n_4$.}
\end{table}
\end{center}

\begin{figure}
\centering
\includegraphics[scale=0.75]{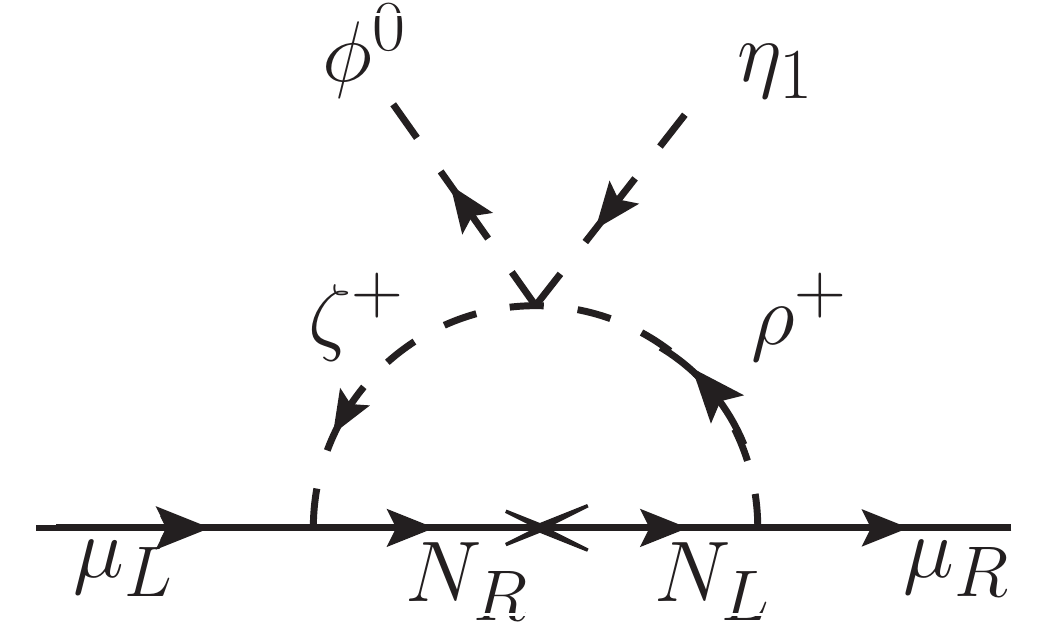}
\caption{One-loop contribution to muon mass.}
\label{fig1}
\end{figure}

Clearly, one can consider additional field content in order to provide a UV complete realisation for such higher dimensional operators for first and second generation masses. For example, muon mass can arise at one-loop level, in scotogenic fashion \cite{Ma:2006km}, after introducing the particles shown in table \ref{table3}. The corresponding one-loop diagram is shown in figure \ref{fig1}.

\begin{center}
\begin{table}
\begin{tabular}{|c|c|c|}
\hline
Particle & $SU(3)_c \times SU(2)_L \times U(1)_Y$ & $U(1)_X$  \\
\hline
$N_{L,R} $ & $(1,1,0)$ & $-\frac{1}{4}$ \\
$\zeta = (\zeta^+, \zeta^0)$ & $(1,2,\frac{1}{2})$ & $-\frac{5}{4}$ \\
$\rho$ & $(1,1,1)$ & $ -\frac{3}{2}$ \\
\hline
\end{tabular}
\caption{Particles responsible for scotogenic muon mass.}
\label{table3}
\end{table}
\end{center}

Similarly, additional fields can be introduced to generate other charged fermion as well as Dirac neutrino masses of first and second generations at radiative level. Here we focus only the new physics responsible for muon mass origin at one-loop in the context of dark matter, muon $(g-2)$ and LHC constraints. The relevant part of the Lagrangian for muon mass is given by
\begin{align}
    \mathcal{L} \supset -y_{\zeta} \bar{L_{\mu}} \tilde{\zeta} N_R - M_N \bar{N_L}N_R - y_{\rho} \bar{N_L} \rho \mu_R -\lambda \Phi \zeta \eta^{\dagger}_1 \rho^{\dagger} +  {\rm h.c.}
\end{align}
As noticed from the above Lagrangian, the newly introduced fields for scotogenic muon mass, always appear in pairs of the form $\psi^{\dagger}_1 \psi$. This is due to the chosen $U(1)_X$ charge assignments of these fields. Therefore, the Lagrangian possesses a global $U(1)_D$ symmetry under which the fields shown in table \ref{table3} can have non-trivial transformations while the SM fields transform trivially \cite{Ma:2021aag}. As none of the scalar fields in table \ref{table3} acquire any vacuum expectation value (VEV), this symmetry remains unbroken, keeping the lightest particle with non-trivial $U(1)_D$ charge stable and hence the DM candidate.

The one-loop muon mass can be estimated as
\begin{equation}
    m_{\mu}=\frac{Y_{\zeta} Y_{\rho}}{16\pi^2} \frac{\lambda v u_1}{2} \frac{M_N}{M_{\chi^+_1} M_{\chi^+_2}} I(x_1, x_2)
\end{equation}
where $M_{\chi^+_1}, M_{\chi^+_2}$ are physical masses of scalars in loop, which can be derived by diagonalising the charged scalar mass matrix given in Appendix \ref{appen1}. Here $v, u_1$ denote the VEV of the neutral component of the SM Higgs doublet $\Phi$ and singlet scalar $\eta_1$ respectively. The physical mass eigenstates arise due to mixing of $\zeta^+, \rho^+$ by angle given by
\begin{equation}
    \sin{2 \theta_{\rm ch}} = \frac{\lambda v u_1}{M^2_{\chi^+_1}-M^2_{\chi^+_2}}.
    \label{eq:mixing1}
\end{equation}
The loop function $I(x_1, x_2)$ is given by
\begin{equation}
    I(x_1, x_2)=\frac{\sqrt{x_1 x_2}}{x_1-x_2} \left ( \frac{x_1}{x_1-1} \ln{x_1}- \frac{x_2}{x_2-1} \ln{x_2} \right)
\end{equation}
where $x_{1}=M^2_{\chi^+_1}/M^2_N$, $x_{2}=M^2_{\chi^+_2}/M^2_N$. The effective coupling of the SM Higgs to muon can be calculated from the same muon mass diagram as 
\begin{equation}
    Y^{\rm eff}_{\mu} = \frac{\sqrt{2} m_{\mu}}{v} \bigg [ \cos^2{(2\theta_{\rm ch})}+\frac{1}{2} \sin^2{(2\theta_{\rm ch})} \frac{\sqrt{x_1 x_2}}{I(x_1, x_2)} \left ( \frac{I(x_1)}{x_1}+\frac{I(x_2)}{x_2} \right ) \bigg ]
    \label{muoncoupling}
\end{equation}
where 
$$ I(x)=\frac{x}{x-1}-\frac{x \ln{x}}{(x-1)^2}. $$
For details of other fermion masses including neutrinos, one may refer to \cite{Ma:2021aag}. The physical scalar spectrum and the couplings are given in Appendix \ref{appen1}. Clearly, the muon coupling to the SM Higgs gets changed from the usual SM value $\sqrt{2} m_{\mu}/v$ to the one shown in equation \eqref{muoncoupling} above. As can be seen from the full scalar potential of the model given in Appendix \ref{appen1}, in addition to $\lambda \Phi \zeta \eta^{\dagger}_1 \rho^{\dagger}$ term discussed above, there exist other quartic couplings of SM Higgs with scalars like $\zeta, \rho$. Since $\zeta, \rho$ also couple to muons, such additional quartic couplings can also lead to anomalous Higgs coupling to muons without contributing to muon mass at one-loop. However, we have considered such additional quartic couplings to be small so that dominant contribution to muon anomalous coupling to Higgs arises from the same quartic coupling which also gives rise to radiative muon mass as discussed above. This anomalous muon coupling to the SM Higgs can be constrained from the LHC observations as we discuss in one of the upcoming sections. While there is no role of singlet scalar $\eta_2$ in muon mass generation at one-loop, it is required to generate other fermion masses within a minimal setup as discussed in \cite{Ma:2021aag}. Considering the VEV of $\eta_2$ to be $u_2$, the mass of $U(1)_X$ gauge boson after symmetry breaking is $M_{Z_X}=g_X \sqrt{u^2_1+9u^2_2}/4$.

\begin{figure}
\centering
\includegraphics[scale=0.75]{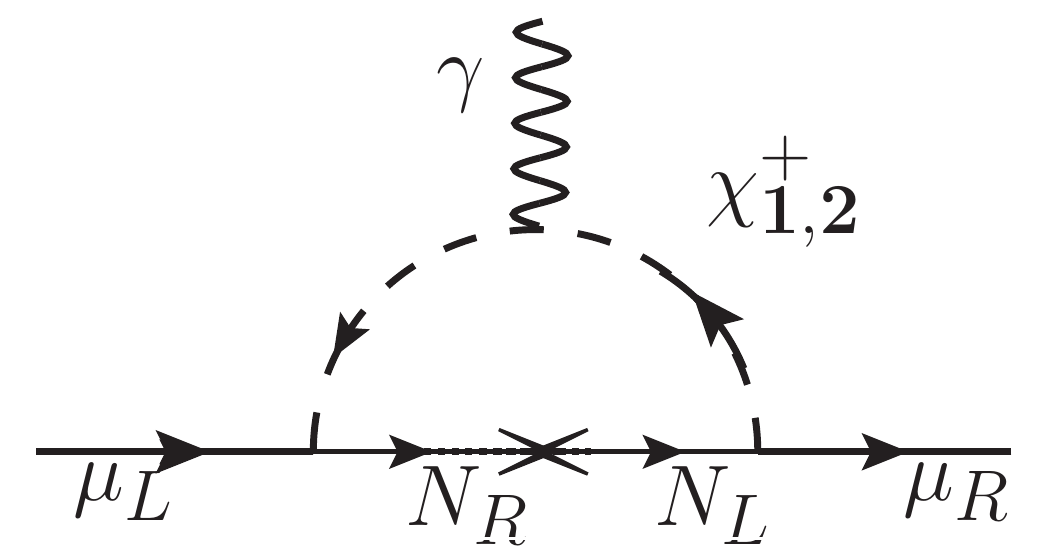}
\caption{One-loop contribution to muon $(g-2)$ from charged scalars.}
\label{fig2}
\end{figure}
\section{Muon Anomalous Magnetic Moment}
\label{sec:g-2}
As mentioned before, there is a $4.2\sigma$ discrepancy between muon AMM predictions of theory and experimental measurements and can potentially be explained with BSM physics. In the $U(1)_X$ gauge  model we discuss here, there are two different contributions to muon $(g-2)$: one from charged scalars in the loop and another where $U(1)_X$ gauge boson goes in the loop. While the contribution from $U(1)_X$ gauge boson loop is sub-dominant for typical TeV scale masses, the contribution from charged scalar loop can be enhanced. This is because, the same loop particles also give rise to muon mass thereby removing the additional loop factor from muon $(g-2)$ contributions \cite{Fraser:2015zed}. Similar discussions on muon $(g-2)$ in radiative muon mass models have also appeared recently in \cite{Baker:2021yli}. The charged scalar loop contribution to muon $(g-2)$ in our model is shown in figure \ref{fig2} where $\chi^{+}_{1,2}$ are the mass eigenstates of $\zeta^+, \rho^+$ after symmetry breaking. The corresponding contribution to muon $(g-2)$ is given by \cite{Baker:2021yli}
\begin{align}
    \Delta a_{\mu} & =\frac{m^2_{\mu}}{M^2_N} \left ( \frac{x_1 \ln{x_1}}{1-x_1}- \frac{x_2 \ln{x_2}}{1-x_2} \right)^{-1} \bigg [ \frac{3x_1-1}{(1-x_1)^2} -\frac{3x_2-1}{(1-x_2)^2} + \frac{2x^2_1 \ln{x_1}}{(1-x_1)^3}- \frac{2x^2_2 \ln{x_2}}{(1-x_2)^3} \nonumber \\
    & + 2 \left ( \frac{1}{1-x_1}-\frac{1}{1-x_2}+ \frac{x_1 \ln{x_1}}{(1-x_1)^2}-\frac{x_2 \ln{x_2}}{(1-x_2)^2} \right ) \bigg ].
\end{align}

\begin{figure}[h]
\centering
\includegraphics[scale=0.75]{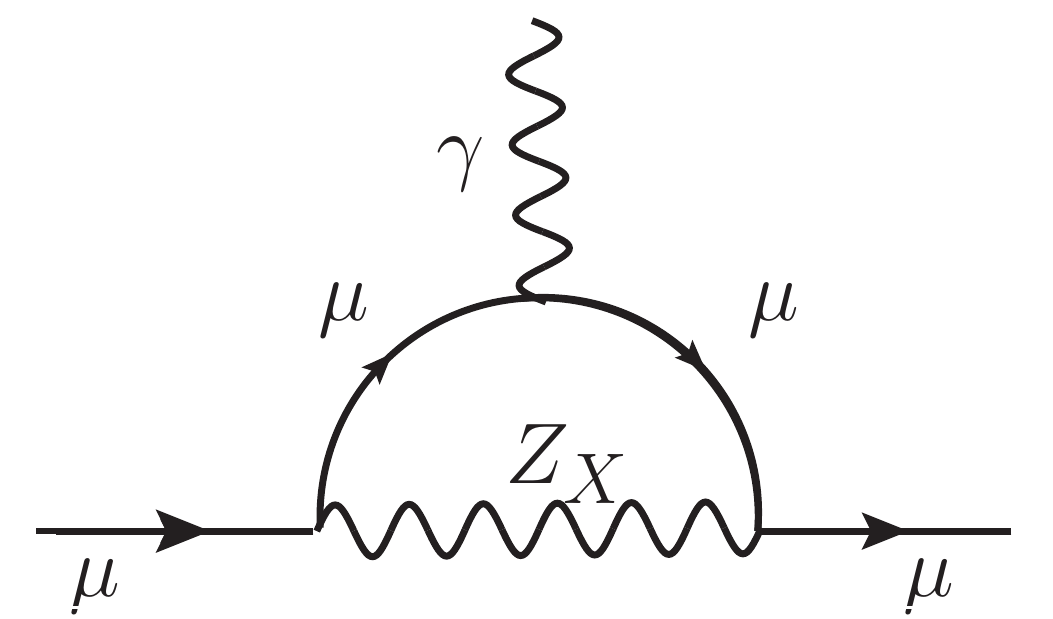}
\caption{One-loop contribution to muon $(g-2)$ due to extra $U(1)$ gauge boson.}
\label{figX}
\end{figure}

The neutral $U(1)_X$ gauge boson contribution to muon AMM (shown in figure \ref{figX}) can be written as \cite{Brodsky:1967sr, Baek:2008nz, Queiroz:2014zfa}
	\begin{equation}
		\Delta a_{\mu} = \frac{\alpha_X}{2\pi} \int^1_0 dx \frac{2m^2_{\mu} x^2 (1-x)}{x^2 m^2_{\mu}+(1-x)M^2_{Z'}} \approx \frac{\alpha_x}{2\pi} \frac{2m^2_{\mu}}{3M^2_{Z'}}
	\end{equation}
	where $\alpha_X=g^2_{X}/(4\pi)$. As shown in earlier works  \cite{Bauer:2018onh, Borah:2020jzi, Borah:2021jzu, Borah:2021mri, Borah:2021khc}, the only allowed region where such neutral gauge boson contribution can explain muon AMM is in the sub-GeV regime with corresponding gauge coupling smaller than $10^{-3}$. Since we consider heavy gauge boson limit, the contribution from such neutral gauge bosons remain suppressed. In fact, since $U(1)_X$ gauge boson couples to electrons as well, the bounds from low energy experiments related to dark photon searches are likely to rule out the low mass regime completely \cite{Bauer:2018onh} leaving us with the explanation of muon AMM from charged scalar loop only.
	
	An important observation about muon g-2 is that if the muon
mass originates at tree level, as in the SM, then a loop contribution from a scalar
and a fermion is positive if the scalar (fermion) is neutral (charged), but
negative if the scalar (fermion) is charged (neutral).  However, if the muon mass is radiative in one-loop coming from a scalar and a fermion as in our model, then the sign reverses. Therefore, even with charged scalar loop shown in figure \ref{fig2}, we can still explain positive $\Delta a_{\mu}$.

\section{Electroweak Precision Constraints}
\label{sec:S&T}
Another constraint on the model parameters can arise due to the electroweak precision data (EWPD) encoded in Peskin-Takeuchi oblique parameters S and T. Due to the presence of new scalar doublet ($\zeta$) and charge singlet scalar ($\rho$), these oblique parameters can receive additional contributions. As shown in \cite{Grimus:2008nb, Cao:2017ffm}, the charged singlet scalar ($\rho$) contributes to the S parameter only and does not affect the T parameter at one loop level. Also, the corresponding contribution of singlet scalar remains small, well within error bars. The contributions due to the scalar doublet ($\zeta$) can be written as\cite{Jueid:2020rek}
\begin{align}
S &= \frac{1}{12\pi}
 \ln \frac{M_{\zeta_R}^2}{M_{\chi_1^+}^2} ,
\\ \nonumber
T &= \frac{1}{16\pi^2 \alpha v^2} F(M_{\chi_1^+}^2,M_{\zeta_R}^2),
\end{align}
where $F(x,y)$ is the loop function and can be expressed as
\begin{align}
F(x,y) =
\left\{
\begin{array}{ll}
\frac{x+y}{2}-\frac{xy}{x-y}\ln \frac{x}{y}, & \hbox{ if } x \neq y;
\\
0, & \hbox{ if } x =y.
\end{array}
\right.
\end{align}
The present best fit values of $S=0.02\pm 0.07$ and $T=0.07\pm0.06$ \cite{ParticleDataGroup:2018ovx} can be used for deriving the constraint on the model parameters as we discuss in upcoming sections. 

\section{Electric dipole moment and lepton flavour violation}
\label{sec:edmlfv}
Similar to the anomalous magnetic moment discussed above, electric dipole moment (EDM) of leptons is a flavour conserving observable which is a measure of the coupling of the lepton's spin to an external electric field. In the SM, lepton EDMs are vanishingly small and hence any experimental observation can be a clear sign of BSM physics. While in the SM, EDM of lepton like muon arises only at four loop level, in the present model, we can have muon EDM at one-loop level itself via a diagram  similar to the one-loop diagrams for muon $(g-2)$. Since one-loop contribution to muon EDM can be sizeable, one can constrain the model parameters from experimental bound \cite{Muong-2:2008ebm}
\begin{equation}
\lvert d_{\mu} \rvert/e < 1.9 \times 10^{-19} \; \text{cm}.
\end{equation}
However, EDM is a CP violating observable and hence depends upon CP violating couplings involved in the one-loop process \cite{Borah:2017leo}. Since rest of our analysis does not rely upon new sources of CP violation, we can tune them appropriately to keep the resulting EDM within experimental limit.

Another flavour observable is related to charged lepton flavour violation (CLFV) like $\mu \rightarrow e \gamma$ which can naturally arise in BSM scenarios like radiative mass models. Experimental constraints on this rare decay process $\text{Br}(\mu \rightarrow e \gamma) < 4.2 \times 10^{-13}$ at $90\%$ confidence level \cite{MEG:2016leq} can be used to constrain the parameter space of such models. In order to realise such flavour violating decays, the particles in the loop need to couple to different generations of fermions. However, due to non-universal $U(1)_X$ charges in our model, the fields responsible for radiative muon mass as well as muon $(g-2)$, do not couple to other lepton generations. Therefore, we do not have such one-loop CLFV processes and hence they do not impose any additional constraints on the parameter space.

\section{Collider Constraints}
\label{sec:lhc}
Collider constraints can primarily apply on SM Higgs decay into muons as the effective coupling is changed in such radiative muon mass models. Additional constraints can apply to physical masses of charged scalars as well as other particles having electroweak interactions from direct search bounds. The modifications in Higgs decay into muons, relative to the SM can be written the corresponding ratio of branching fractions as
\begin{equation}
0.8 \times 10^{-4} < {\rm BR} \left(h \rightarrow \mu^+ \mu^- \right) < 4.5 \times 10^{-4}
\end{equation}
as given by the CMS collaboration \cite{Sirunyan:2020two}. Similar bound has been reported by the ATLAS collaboration \cite{Aad:2020xfq} as well.

Higgs to diphoton rate in the model including SM contribution and new charged scalars $\chi^+_{1,2}$ is given by\cite{Djouadi:2005gi}
\begin{equation}
    \Gamma (h \rightarrow \gamma \gamma) = \frac{G_F \alpha^2 m^3_h}{128 \sqrt{2} \pi^3} \bigg \lvert \sum_f N_c Q^2_f A^h_{1/2} (\tau_f) + A^h_1 (\tau_w) + \sum_i g_{hii} Q^2_i A^h_0 (\tau_i) \bigg \rvert^2
    \label{eq:hgg1}
\end{equation}
where $G_F$ is Fermi coupling constant, $\alpha$ is fine structure constant, $N_c$ is the color factor of 
charged fermion in loop, $Q_{f,i}$ are electromagnetic charges of fermions and scalars in loop and $\tau_i  
= m^2_h/4m^2_i$ with $i$ running over all charged particles in loop. The form factors for fermion, vector boson and scalars are given by
$$ A^h_{1/2} (\tau) = 2[ \tau + (\tau-1)f(\tau) ] \tau^{-2}, $$
$$ A^h_1 (\tau) = -[2\tau^2 + 3\tau + 3(2\tau-1) f(\tau) ] \tau^{-2}, $$
$$ A^h_0 (\tau) = -[ \tau -f(\tau) ] \tau^{-2}. $$
The function $f(\tau)$ is given by
\[ f(\tau) = 
\begin{cases}
\text{arcsin}^2 \sqrt{\tau}, & \tau \leq 1 \\
-\frac{1}{4} \left ( \log \frac{1+\sqrt{1-\tau^{-1}}}{1-\sqrt{1-\tau^{-1}}} -i\pi \right )^2, & \tau >1.
\end{cases}
\]
The parameter $g_{hij}$ denotes SM Higgs coupling with the charged scalar $\chi^+_{i} \chi^-_{j}$. They are given by :
\[g_{h11}= - \lambda u_1 \cos{\theta_{\rm ch}} \sin{\theta_{\rm ch}}, \; g_{h22}=  \lambda u_1 \cos{\theta_{\rm ch}} \sin{\theta_{\rm ch}},\; g_{h12}=  \lambda u_1 (\cos^{2}{\theta_{\rm ch}} -\sin^{2}{\theta_{\rm ch}} )\]
where $\theta_{\rm ch}$ is the mixing angle for $\zeta^+$ and $\rho^+$ as given by Eq. \eqref{eq:mixing1}.
The first two couplings are relevant for $h \rightarrow \gamma \gamma$. The first two terms in Eq. \eqref{eq:hgg1} are due to SM contributions and the last term is due to charged scalars in extra $U(1)_X$ gauge model. So new contributions to $\Gamma (h \rightarrow \gamma \gamma)$ come from the last term and its interference with SM terms.

According to the the latest CMS results \cite{Sirunyan:2021ybb}, the constraints on Higgs to diphoton ratio is $\frac{{\rm BR}(h \rightarrow \gamma \gamma)_{\rm expt}}{{\rm BR}(h \rightarrow \gamma \gamma)_{\rm SM}} = 1.12 \pm 0.09 $ which implies the new contribution should satisfy the constraint
\begin{equation}
    \frac{{\rm BR}(h \rightarrow \gamma \gamma)_{\rm New}}{{\rm BR}(h \rightarrow \gamma \gamma)_{\rm expt}} = 0.0291 \;\; {\rm to}\;\; 0.1735
\end{equation}

Similarly, collider bounds exist on neutral gauge boson mass and corresponding gauge couplings.  The limits from LEP II data constrains such additional gauge sector by imposing a lower bound on the ratio of new gauge boson mass to the new gauge coupling $M_{Z_X}/g_X \geq 7$ TeV \cite{Carena:2004xs, Cacciapaglia:2006pk}. The bounds from ongoing LHC experiment have already surpassed the LEP II bounds. In particular, search for high mass dilepton resonances have put strict bounds on such additional gauge sector coupling to all generations of leptons and quarks with coupling similar to electroweak ones. The latest bounds from the ATLAS experiment \cite{Aaboud:2017buh, Aad:2019fac} and the CMS experiment \cite{Sirunyan:2018exx} at the LHC rule out such gauge boson masses below 4-5 TeV from analysis of 13 TeV data. Such bounds get weaker, if the corresponding gauge couplings are weaker \cite{Aaboud:2017buh} than the electroweak gauge couplings. Also, if the $Z'$ gauge boson couples only to the third generation of leptons, all such collider bounds become much weaker, as explored in the context of DM and collider searches in a recent work \cite{Barman:2019aku}. Similarly, additional scalar sector can also be constrained from collider data. While there are no dedicated LHC searches for singlet charged scalar (like $\rho$ in our model) yet, theoretical studies like \cite{Alcaide:2019kdr} show high luminosity LHC sensitivity upto 500 GeV. For electroweak doublet like $\zeta$, LEP II bounds rule out some part of the parameter space below 100 GeV \cite{Lundstrom:2008ai}. At colliders, if they are produced, they can decay into DM (missing energy) as well as charged leptons (say, muon). Such leptonic final states with missing energy have been studied in several earlier works \cite{Miao:2010rg, Gustafsson:2012aj, Datta:2016nfz}. As a conservative lower limit, we consider all such BSM scalars to be heavier than 100 GeV in our numerical analysis.

\section{Dark Matter}
\label{sec:DM}

The neutral singlet vector like fermion $N_{L, R}$ is the dark matter candidate in this model. Although neutral component of the scalar doublet $\zeta$ could also be a DM candidate, it turns out that the neutral components of $\zeta$ are degenerate leading to a large $Z$ boson mediated DM-nucleon scattering, ruled out by experiments like XENON1T \cite{Aprile:2018dbl}. The situation is similar to sneutrino DM in minimal supersymmetric standard model (MSSM) \cite{Arina:2007tm}. This leaves us with the only choice of fermion singlet being the DM candidate. Since it does not interact with any singlet scalar, so DM phenomenology is dictated by its annihilation via $U(1)_X$ gauge boson only. While for such pure gauge mediated annihilations, the relic is likely to be satisfied near the resonance region $M_{\rm DM} \approx M_{Z_X}/2$, for small mass splitting between DM and charged scalars $\chi^+_{1,2}$ one can have interesting coannihilation effects which depends upon Yukawa couplings dictating both muon mass and $(g-2)$.

The relic abundance of a dark matter particle $\rm DM$, which was in thermal equilibrium at some earlier epoch can be calculated by solving the Boltzmann equation
\begin{equation}
\frac{dn_{\rm DM}}{dt}+3Hn_{\rm DM} = -\langle \sigma v \rangle (n^2_{\rm DM} -(n^{\rm eq}_{\rm DM})^2)
\label{eq:dmbe1}
\end{equation}
where $n_{\rm DM}$ is the number density of the dark matter particle $\rm DM$ and $n^{\rm eq}_{\rm DM}$ is the number density when $\rm DM$ was in thermal equilibrium. $H$ is the Hubble expansion rate of the Universe and $ \langle \sigma v \rangle $ is the thermally averaged annihilation cross section of the dark matter particle $\rm DM$. In terms of partial wave expansion $ \langle \sigma v \rangle = a +b v^2$. Numerical solution of the Boltzmann equation above gives \cite{Kolb:1990vq,Scherrer:1985zt}
\begin{equation}
\Omega_{\rm DM} h^2 \approx \frac{1.04 \times 10^9 x_F}{M_{\text{Pl}} \sqrt{g_*} (a+3b/x_F)}
\end{equation}
where $x_F = M_{\rm DM}/T_F$, $T_F$ is the freeze-out temperature, $M_{\rm DM}$ is the mass of dark matter, $g_*$ is the number of relativistic degrees of freedom at the time of freeze-out and and $M_{\text{Pl}} \approx 2.4\times 10^{18}$ GeV is the Planck mass. Dark matter particles with electroweak scale mass and couplings freeze out at temperatures approximately in the range $x_F \approx 20-30$. More generally, $x_F$ can be calculated from the relation 
\begin{equation}
x_F = \ln \frac{0.038gM_{\text{Pl}}M_{\rm DM}<\sigma v>}{g_*^{1/2}x_F^{1/2}}
\label{xf}
\end{equation}
which can be derived from the equality condition of DM interaction rate $\Gamma = n_{\rm DM} \langle \sigma v \rangle$ with the rate of expansion of the Universe $H \approx g^{1/2}_*\frac{T^2}{M_{Pl}}$. 
The thermal averaged annihilation cross section $\langle \sigma v \rangle$ used in Boltzmann equation of \eqref{eq:dmbe1} is given by \cite{Gondolo:1990dk}
\begin{equation}
\langle \sigma v \rangle = \frac{1}{8M^4_{\rm DM}T K^2_2(M_{\rm DM}/T)} \int^{\infty}_{4M^2_{\rm DM}}\sigma (s-4M^2_{\rm DM})\surd{s}K_1(\surd{s}/T) ds
\end{equation}
where $K_i$'s are modified Bessel functions of order $i$, $m$ is the mass of Dark Matter particle and $T$ is the temperature.

If there exists some additional particles having mass difference close to that of DM, then they can be thermally accessible during the epoch of DM freeze out. This can give rise to additional channels through which DM can coannihilate with such additional particles and produce SM particles in the final states. This type of coannihilation effects on dark matter relic abundance were studied by several authors in \cite{Griest:1990kh, Edsjo:1997bg, Bell:2013wua}. As we will see while incorporating all relevant constraints, there exist regions of parameter space where DM fermion can have small mass splitting with charged scalars leading to a region of strong coannihilations. Since the corresponding Yukawa couplings are also required to be large to satisfy other bounds, such coannihilations can in fact lead to suppressed relic abundance. We use the package \texttt{micrOMEGAs} \cite{Belanger:2013oya} to calculate DM relic abundance in the most general way and use \texttt{FeynRules} \cite{Alloul:2013bka} package to prepare the required model files.



\section{Results and Conclusion}
\label{sec:conclude}

\begin{figure}[h!]
 \includegraphics[scale=0.5]{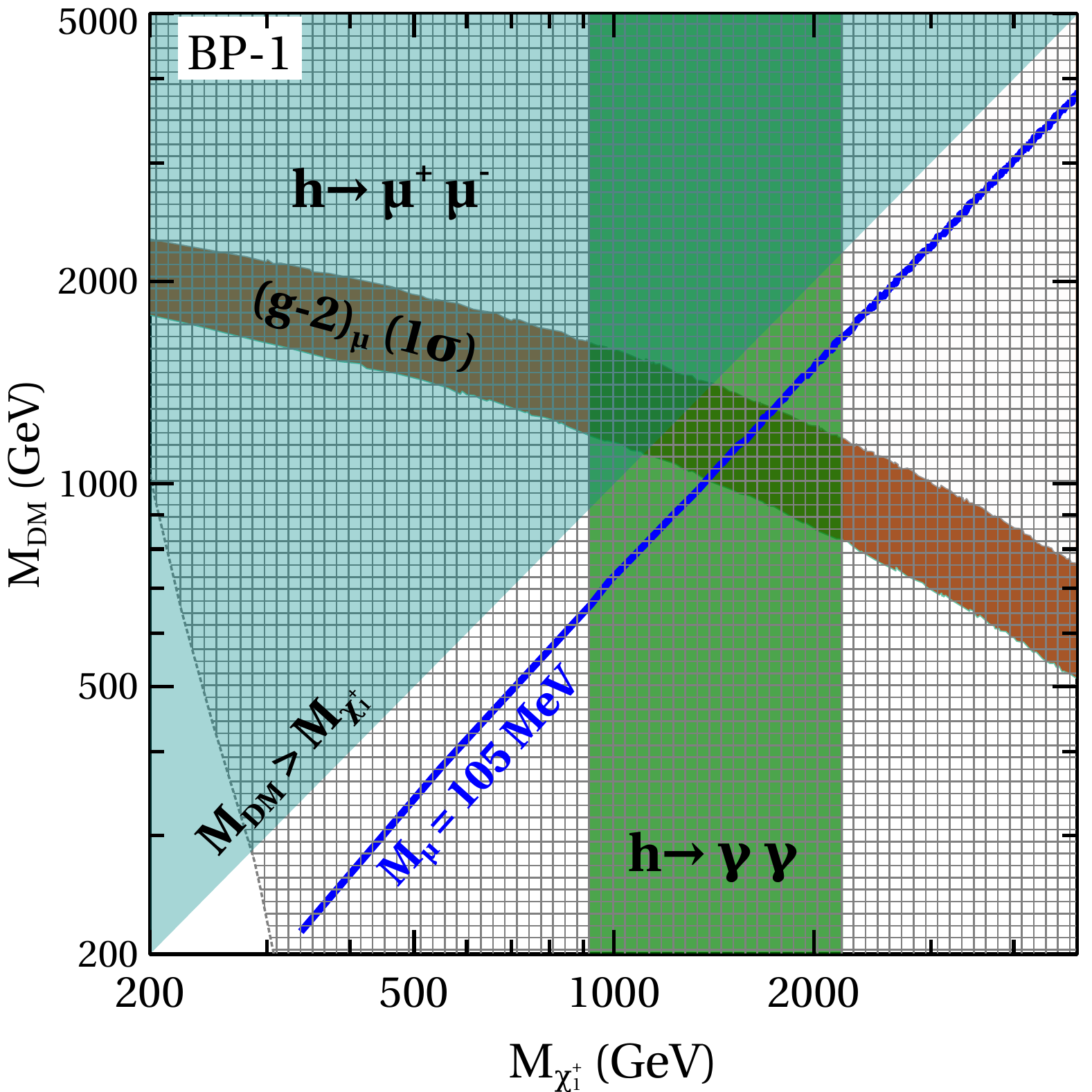}
\caption{Common parameter space satisfying Muon mass (blue line), Muon $(g-2)$ (the brown band), $h\rightarrow \gamma \gamma$ (vertical green band) and $h\rightarrow \mu^+ \mu^-$ (the grey mesh) for a chosen benchmark BP-1.}
\label{fig:BP1}
\end{figure}

Let us now discuss all possible phenomenological consequences of our model. We will consider the constraints coming form the AMM of muon, muon mass, the decay of SM Higgs to $\gamma \gamma$ and $\mu^+ \mu^-$, and finally the relic abundance of DM respectively. The important parameters for these different observable are the following: 
$${\rm M_{\chi_1^+}, M_{\zeta_R}, M_{DM}, \theta_{ch}, y_{\zeta}=y_{\rho}, \lambda,g_X, M_{Z_X}}.$$ 
However, all the other observable are independent of $g_X$ and $M_{Z_X}$ except the relic density of DM and we will discuss the role of these parameters first. In figure \ref{fig:BP1}, we have shown the allowed parameter space in $M_{\rm DM}$ vs $M_{\chi_1^+}$ plane, from muon mass (blue line), muon $(g-2)$ (the brown band), $h\rightarrow \gamma \gamma$ (vertical green band) and 
$h\rightarrow \mu^+ \mu^-$ (the grey mesh). We have fixed all the other parameters according to the BP-1 shown is table \ref{tab:BP} and one can clearly see that all these different regions coincide with each other in a very tiny region in $M_{\rm DM}$ vs $M_{\chi_1^+}$ plane.

\begin{figure}[h!]
\includegraphics[scale=0.5]{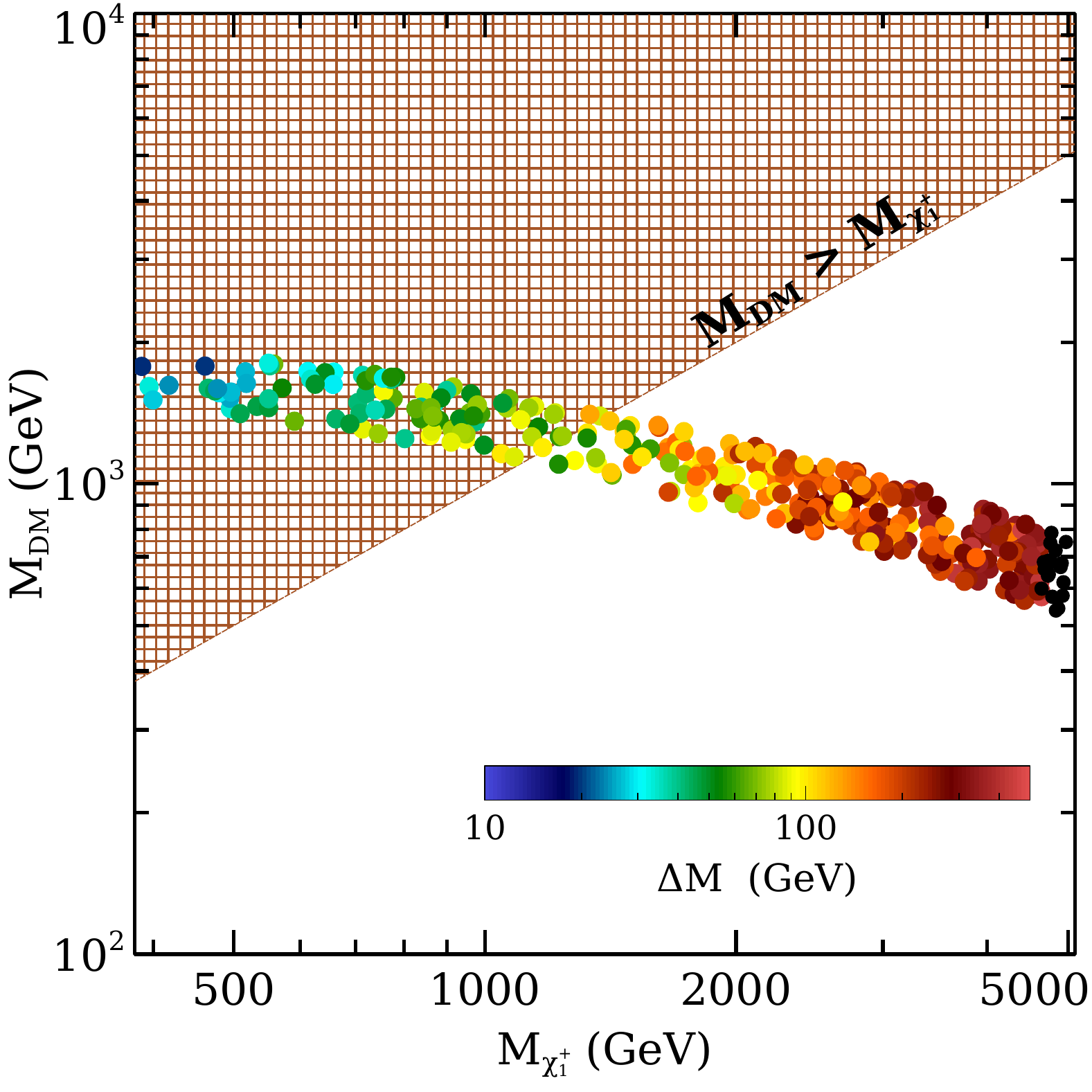}\, \includegraphics[scale=0.5]{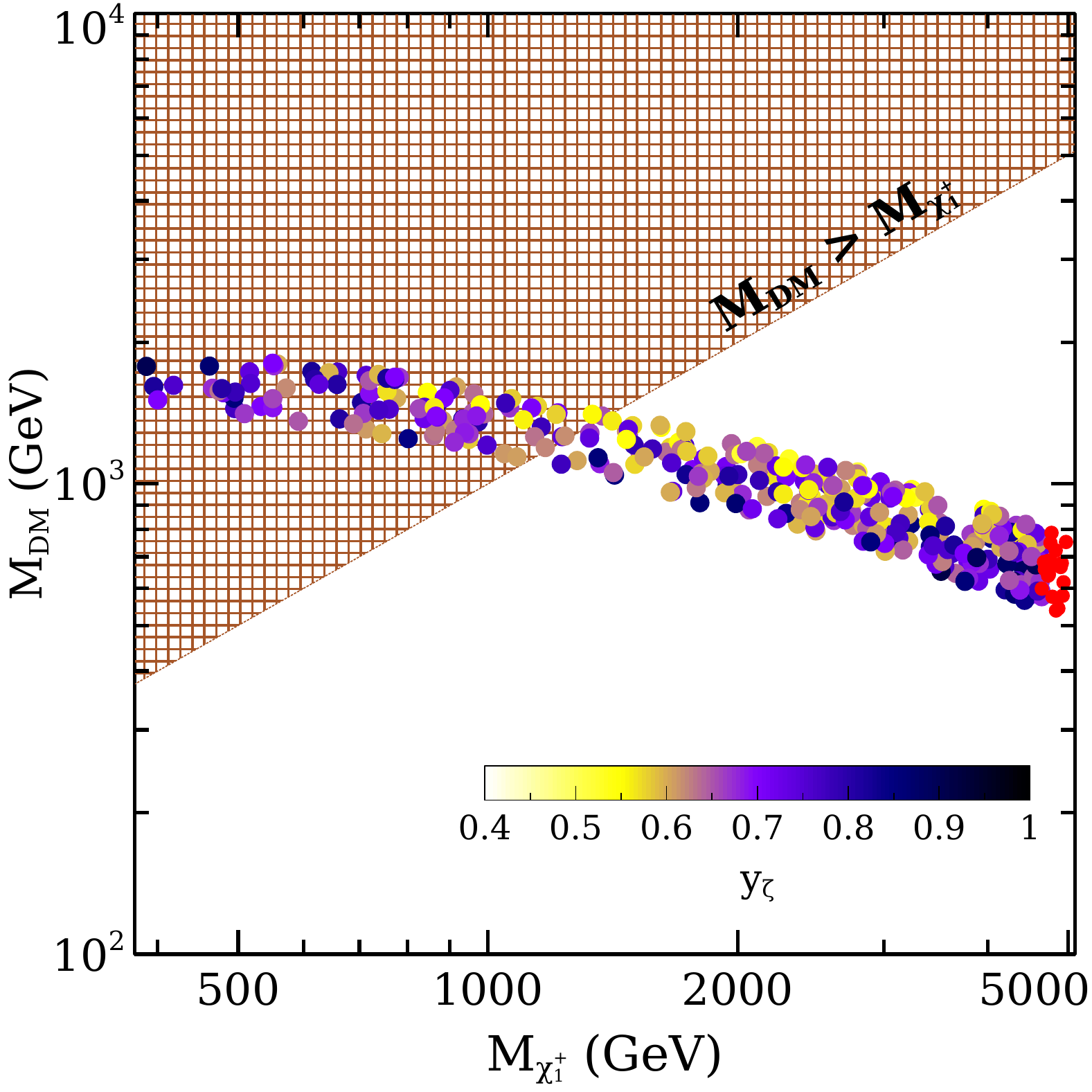}
\caption{Allowed parameter space from all relevant constraints satisfying muon mass, muon $(g-2)$, $h\rightarrow \gamma \gamma$  and $h\rightarrow \mu^+ \mu^-$. The color code represents the mass splitting between the $M_{\chi_1^+}$ and $M_{\zeta_R}$ in left panel whereas the color code shows the variation of the Yukawa coupling $y_{\zeta}$ in the right panel.}
\label{fig:scan:1}
\end{figure}

In figure \ref{fig:scan:1}, we have shown the allowed parameter space in the same $M_{\rm DM}$ vs $M_{\chi_1^+}$ plane by varying all the other parameters as mentioned in table \ref{tab:scan}. In the left panel, the color code represents the mass splitting between the $M_{\chi_1^+}$ and $M_{\zeta_R}$ where as in the right panel, the color code shows the variation of the Yukawa coupling $y_{\zeta}$.  For simplicity, we have assumed equality of Yukawa couplings $y_{\zeta}=y_{\rho}$. Any deviation from this equality is unlikely to bring substantial change in our results. In spite of the presence of many different parameters, a very small region of parameter space is allowed from all the above-mentioned constraints. One can also note that we require quite large $y_{\zeta}$ ($<0.5$) to satisfy all possible constraints. Finally, we have shown the constraints coming from the electroweak precision observable as discussed in section \ref{sec:S&T}. A very small region of the parameter space is excluded from the EWPD constraints\footnote{Note that we have made a conservative estimate by considering a pure scalar doublet contribution. The actual estimate will involve both doublet and singlet scalar contributions with possible interference, a full calculation of which is beyond the scope of present work.} as shown in the black and red coloured points in the high mass regime of charged scalar in figure \ref{fig:scan:1}. While the band consisting of coloured points satisfy all relevant bounds, the upper half of the plane (shaded) is disfavoured as it corresponds to unstable DM candidate.

\begin{center}
{\tiny \begin{table}
\begin{tabular}{|p{1.5cm} |p{2.1cm}|p{2.1cm}|p{2.1cm}|p{1.3cm}| p{1.3cm}| p{1.3cm}|p{1.2cm}|p{2.0cm}|}
 \hline
 \multicolumn{9}{|c|}{Benchmark points} \\
 \hline
\ \ \ \  &\ $M_{\chi_1^+}$ (GeV)&\ $M_{\zeta_R}$ (GeV) & $M_{\rm DM}$ (GeV) & $\sin \theta_{\rm ch} $&  $y_{\zeta}=y_{\rho}$&\ \ \ \ \ $\lambda$ & $g_X$ & $M_{Z_{X}}$ (GeV)\\
 \hline
 \ \ \ BP-1  & 200-3000 & $M_{\chi_1^+}-$71.43 & 200-5000 & 0.8741 & 0.6756 & -0.8327 & \ \ \ \ $-$ &\ \ \ \ \ \ \ $-$ \\
 \hline
 \ \ \ BP-1/2  & $M_{DM}$ + 15 & $M_{DM}$ + 10 & 500-3000 & 0.887 & 0.792 & -0.862 & 0.009 & \ \ \ \ 2813\\
 \hline
  \ \ \ BP-2/2  & $M_{DM}$ + 105 & $M_{DM}$ + 100 & 500-3000 & 0.887 & 0.792 & -0.862 & 0.009 & \ \ \ \ 2813\\
 \hline
 \ \ \ BP-3/2  & $M_{DM}$ + 255 & $M_{DM}$ + 250 & 500-3000 & 0.887 & 0.792 & -0.862 & 0.009 & \ \ \ \ 2813\\
 \hline
  \ \ \ BP-1/3  & $M_{DM}$ + 15 & $M_{DM}$ + 10 & 500-3000 & 0.9156 & 0.644 & -0.282 & 0.038 & \ \ \ \ 2447\\
 \hline
   \ \ \ BP-2/3  & $M_{DM}$ + 105 & $M_{DM}$ + 100 & 500-3000 & 0.9156 & 0.644 & -0.282 & 0.038 & \ \ \ \  2447\\
 \hline
  \ \ \ BP-3/3  & $M_{DM}$ + 605 & $M_{DM}$ + 600 & 500-3000 & 0.9156 & 0.644 & -0.282 & 0.038 &  \ \  \ \ 2447 \\
 \hline\end{tabular} 
\caption{Benchmark points used in numerical analysis.}
\label{tab:BP}
\end{table}}
\end{center}

\begin{table}
\begin{center}
\begin{tabular}{|c|c|}
\hline
Parameters & Range   \\
\hline
\hline
$\rm{M_{\chi_1^+}}$ &  (100 GeV, 5 TeV)\\
$\rm{M_{\chi_1^+}- M_{\zeta_R}}$ &  (1 GeV, 500 GeV)\\
$\rm{M_{DM}}$ &  (1 GeV, 10 TeV)\\
$\rm{\sin \theta_{\rm ch}}$ &  (0.01, 1)\\
$\rm{y_{\zeta}=y_{\rho}}$ &  (0.01, $\sqrt{4\pi}$)\\
$\rm{\lambda}$ &  (-0.001, -1)\\
\hline
\end{tabular}
\end{center}
\caption{The parameters of our model and ranges used in the scan leading to figure \ref{fig:scan:1}.}
\label{tab:scan}
\end{table}

\begin{figure}[h!]
\includegraphics[scale=0.45]{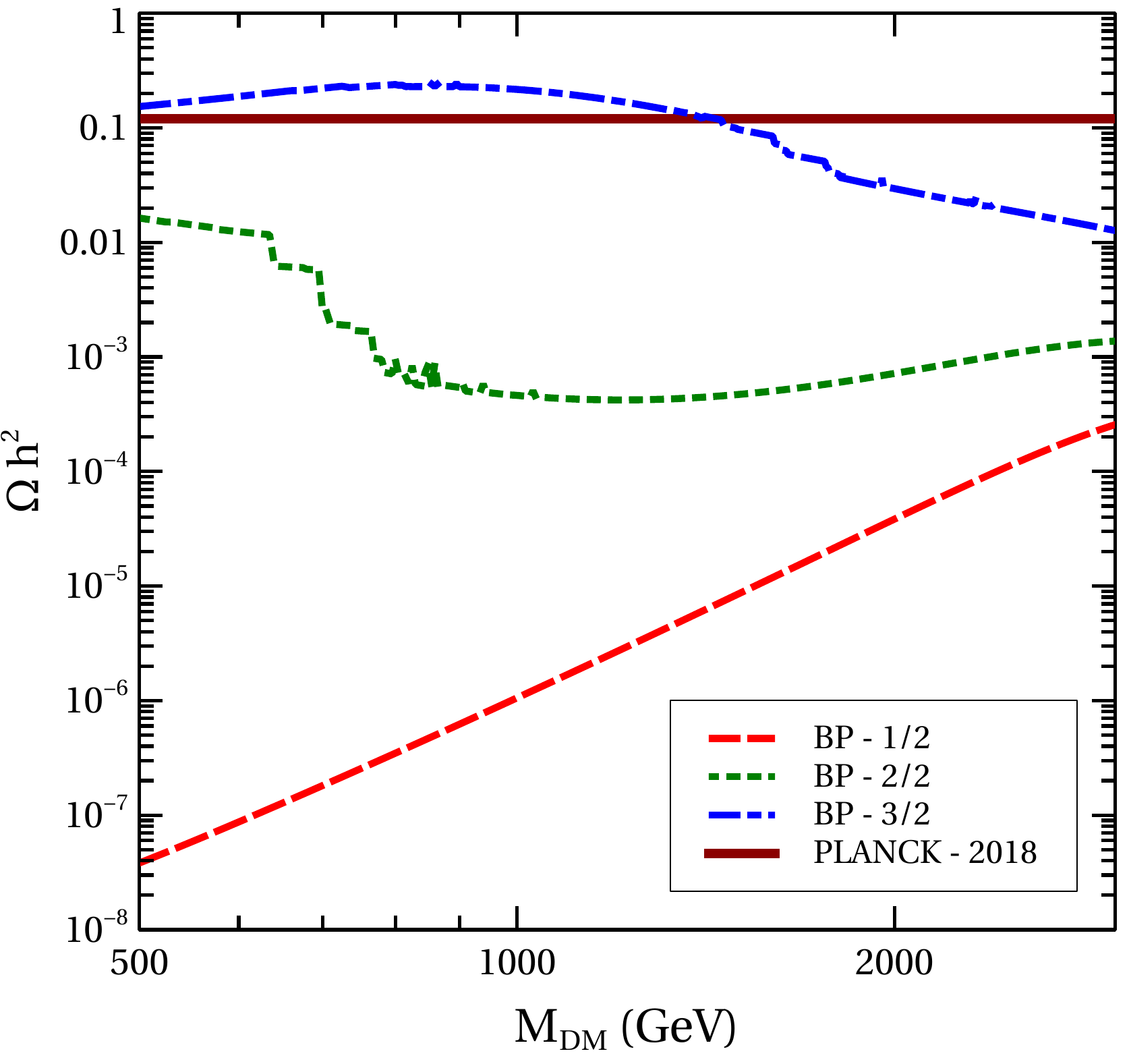}\, \includegraphics[scale=0.45]{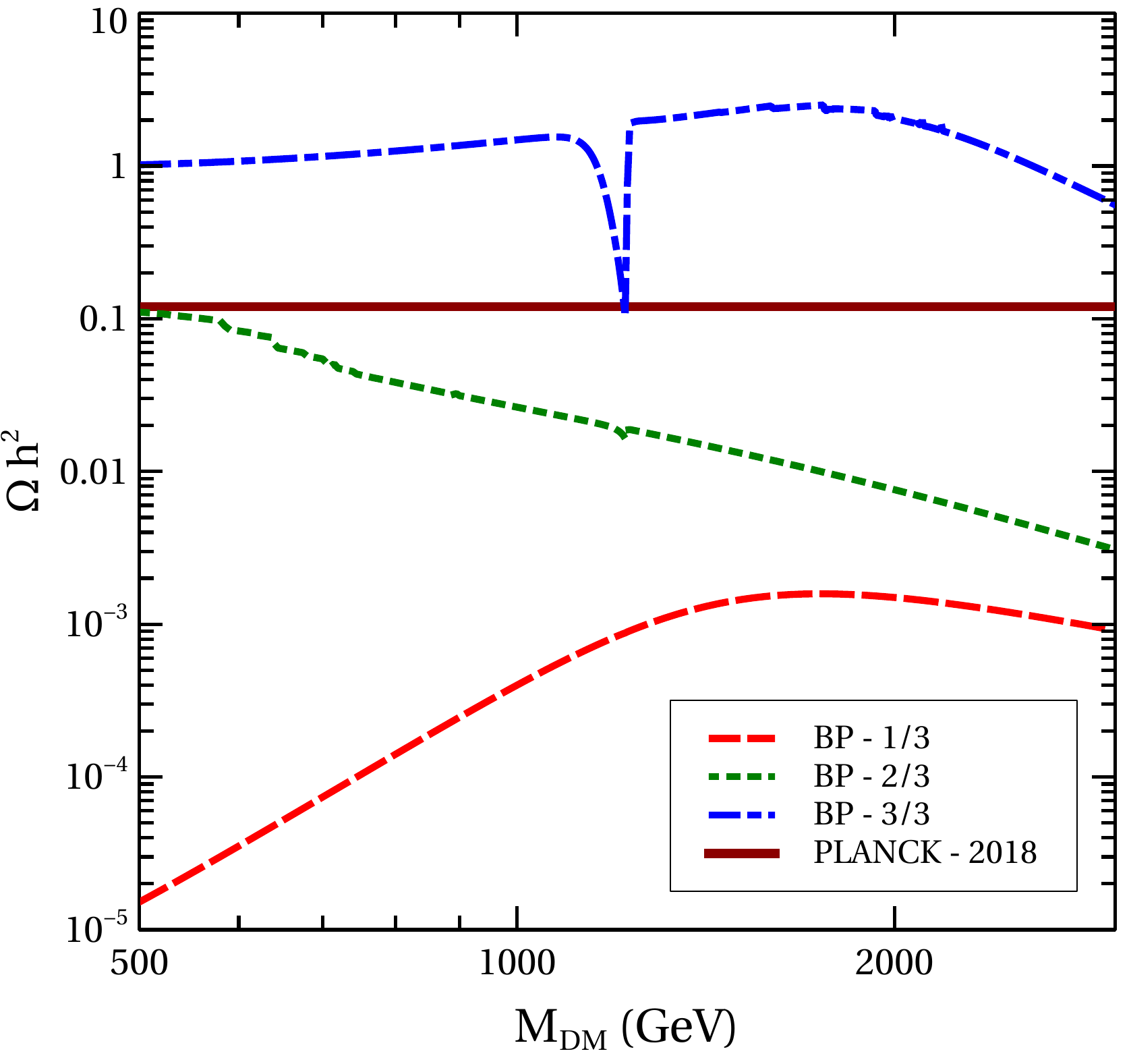}
\caption{The variation of relic abundance of DM as a function of its mass for different benchmark values of other relevant parameters.}
\label{fig:relic}
\end{figure}

\begin{figure}
\includegraphics[scale=0.55]{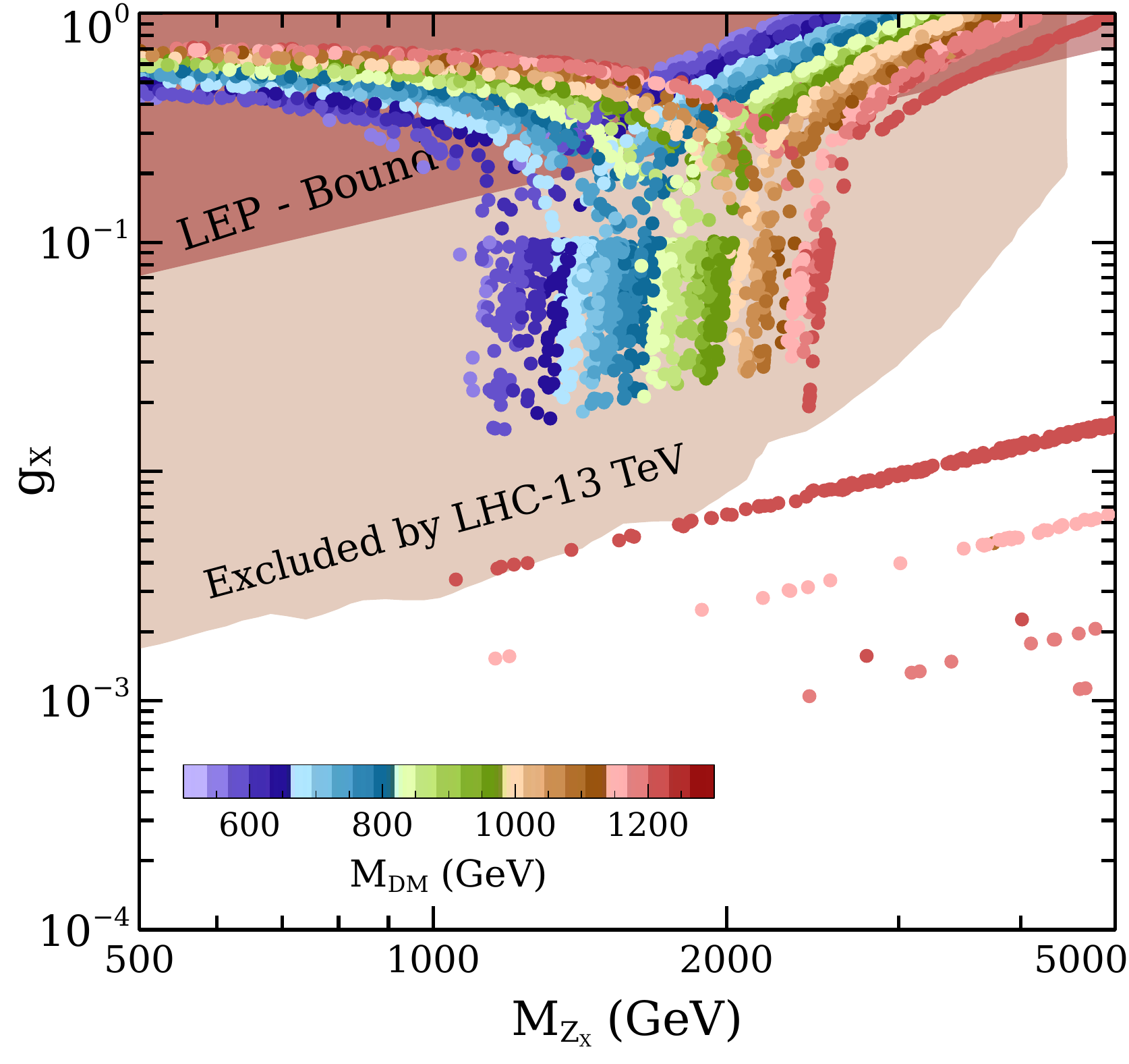} 
\caption{Parameter space in $g_X$ versus $M_{Z_{X}}$ plane favoured from dark matter phenomenology related to relic abundance and direct detection cross section. Dark matter mass range as well as other parameters correspond to allowed points in figure \ref{fig:scan:1} after incorporating other relevant constraints.}
\label{fig:bound-DM}
\end{figure}

So far, we have not taken into account the constraints coming from the observed relic density of DM. As discussed earlier, the DM particles freeze-out from the thermal bath due to the annihilation and co-annihilation processes through the new Yukawa as well as gauge interactions. Figure \ref{fig:relic} represents the relic abundance of DM as a function of its mass ($M_{\rm DM}$) and all the other parameters have been kept fixed according to the benchmark points shown is table \ref{tab:BP}. We have chosen these benchmark points from allowed region shown in figure \ref{fig:scan:1} so that all other constraints are satisfied. The left panel is for very small $g_X \sim 0.009$ whereas the right panel is for slightly larger $g_X \sim 0.03$. One can clearly notice the absence (presence) of $Z_X$ resonance in the left (right) panel due to the smallness (largeness) of the gauge coupling $g_X$. Finally, we have shown the role of both $g_X$ and $M_{Z_X}$ in figure \ref{fig:bound-DM}. Here, we have shown the allowed parameter space in $g_X$ versus $M_{Z_{X}}$ plane, while other parameters are kept fixed at benchmark points allowed from all possible experimental constraints. We consider the allowed points for DM masses as shown in figure \ref{fig:scan:1} and then vary $(g_X, M_{Z_X})$ randomly in the range shown in figure \ref{fig:bound-DM}. The scattered points in figure \ref{fig:bound-DM} correspond to DM masses (shown in colour bar) which satisfy correct relic abundance. The effect of DM annihilation mediated by $Z_X$ is clearly visible for $Z_X$ masses close to resonance regime while the points away from resonance will satisfy relic due either due to large gauge coupling $g_X$ or coannihilation with scalars. The grey shaded region in figure \ref{fig:bound-DM} corresponds to the exclusion limits from the LHC searches for heavy resonances decaying into lepton pairs \cite{Aaboud:2017buh, Aad:2019fac, Sirunyan:2018exx}. The brown shaded region corresponds to the LEP bound $M_{Z_X}/g_X \geq 7$ TeV. The most important point to note here is that the LHC-13 TeV data excludes a broad region of parameter space of our model. In order to implement the LHC bound we compute the dilepton production cross section at 13 TeV center of mass energy at the LHC using the package {\tt MADGRAPH}~\cite{Alwall:2014hca}. As the first generation quarks have $U(1)_X$ charges more than unity, we get a stricter bound on $g_X-M_{Z_X}$ parameter space compared to other universal Abelian extensions like gauged $B-L$. Clearly, the benchmark points shown in the last three rows of table \ref{tab:BP} are already disallowed by the LHC bounds. In fact, only a handful of DM masses from Fig. \ref{fig:scan:1} survive LHC bounds, as shown in Fig. \ref{fig:bound-DM}. Due to constant DM mass but varying $g_X, M_{Z_X}$, many of these points seem to fall on a line in the allowed region of Fig. \ref{fig:bound-DM}. With a much bigger scan size, the allowed region can be filled with more points allowed from all relevant constraints. Clearly, all these allowed points correspond to small values of gauge coupling $g_X$ and hence coannihilation effects play dominant role in generating correct DM relic. The mass splitting of DM and scalars are in the range of 150-500 GeV while the corresponding Yukawa couplings are of order unity leading to efficient coannihilations for DM masses in 1200-1500 GeV range falling in the allowed region. We also check that the points allowed by LHC bounds are also allowed from DM direct detection bounds from XENON1T experiment \cite{Aprile:2018dbl}

To summarise, we have studied an Abelian gauge extension of the standard model with radiative muon mass leading to anomalous magnetic moment as well as anomalous Higgs coupling of muon having very interesting consequences at experiments. While a positive muon $(g-2)$ has been reported recently by the Fermilab experiment confirming the Brookhaven measurements made much earlier, the anomalous Higgs coupling to muon can be probed at the LHC. The model also predicts a stable fermion singlet dark matter candidate which goes inside radiative muon mass loop in scotogenic fashion. Taking into account of all relevant constraints related to muon mass along with its $(g-2)$, Higgs coupling to muons, Higgs to diphoton decay, direct search bounds from colliders as well as dark matter phenomenology lead to a tiny region of parameter space that can be probed at future experiments. 

\acknowledgments
DB acknowledges the support from Early Career Research Award from the Science and Engineering Research Board (SERB), Department of Science and Technology (DST), Government of India (reference number: ECR/2017/001873). DN would like to thank Dr. Najimuddin Khan for fruitful discussions related to collider bounds.

\appendix
\section{Scalar mass spectrum and couplings}
\label{appen1}
The complete scalar potential of the model can be written as
\begin{eqnarray}\nonumber
V(\phi, \eta_1,\eta_2,\zeta,\rho)&=& -\mu_{\phi}^2\,\left(\Phi^\dagger\Phi\right) + \lambda_\phi\,\left(\Phi^\dagger\Phi\right)^2 + \mu_{\zeta}^2\,\left(\zeta^\dagger\zeta\right) +\lambda_{\zeta}\, \left(\zeta^\dagger\zeta\right)^2 -\mu_{\eta_1}^2 \left(\eta_1^\dagger\eta_1\right)+ \lambda_{\eta_1}\left(\eta_1^\dagger\eta_1\right)^2 \\ \nonumber
&&  -\mu_{\eta_2}^2 \left(\eta_2^\dagger\eta_2\right)+\lambda_{\eta_2} \left(\eta_2^\dagger\eta_2\right)^2 +\mu_{\rho}^2 \left(\rho^\dagger\rho\right)+ \lambda_{rho} \left(\rho^\dagger\rho\right)^2+ \lambda_{\phi\zeta}\left(\Phi^\dagger\Phi\right) \left(\zeta^\dagger\zeta\right) \\ \nonumber
&&   + \lambda_{\phi\eta_1}\left(\Phi^\dagger\Phi\right) \left(\eta_1^\dagger\eta_1\right) + \lambda_{\phi\eta_2} \left(\Phi^\dagger\Phi\right) \left(\eta_2^\dagger\eta_2\right) + \lambda_{\phi\rho} \left(\Phi^\dagger\Phi\right) \left(\rho^\dagger\rho\right) + \lambda_{\zeta\eta_1} \left(\zeta^\dagger\zeta\right)  \left(\eta_1^\dagger\eta_1\right) \\ \nonumber
&&  +\lambda_{\zeta\eta_2} \left(\zeta^\dagger\zeta\right)  \left(\eta_2^\dagger\eta_2\right)+ \lambda_{\zeta\rho} \left(\zeta^\dagger\zeta\right)  \left(\rho^\dagger\rho\right)+ \lambda_{\eta_1\eta_2}\left(\eta_1^\dagger\eta_1\right)\left(\eta_2^\dagger\eta_2\right)+\lambda_{\eta_1\rho}\left(\eta_1^\dagger\eta_1\right)\left(\rho^\dagger\rho\right)\\
&&  + \lambda_{\eta_2\rho}\left(\eta_2^\dagger\eta_2\right)\left(\rho^\dagger\rho\right)+ \lambda_{\eta_1\eta_2}^\prime \left[ \left(\eta_2^\dagger \eta_1^3\right)+{\rm h.c.}\right]+\lambda \left[\left(\epsilon_{ab}\phi_a \zeta_b \rho^\dagger \eta_1^\dagger\right)+ {\rm h.c.}\right]
\end{eqnarray}
In the last line of the above potential $\epsilon_{ab}$ is an anti-symmetric tensor and $\phi_a, \zeta_b$ are components of the doublet scalars $\Phi, \zeta$ respectively. As mentioned earlier, only the neutral components of $\Phi$ and $\eta_{1,2}$ acquire VEVs, denoted by $v$, $u_{1,2}$ respectively, leading to spontaneous breaking of SM and $U(1)_X$ gauge symmetries respectively.

The minimisation conditions are
\begin{eqnarray}
\mu_\phi^2&=&\frac{1}{2} \left(\lambda_{\phi\eta_1} u_1^2+ \lambda_{\phi\eta_2} u_2^2 + 2 \lambda_\phi v^2\right),\\
\mu_{\eta_1}^2&=&\frac{1}{2} \left(2\lambda_{\eta_1} u_1^2+3 \lambda_{\eta_1\eta_2}^\prime u_1 u_2+ \lambda_{\eta_1\eta_2} u_2^2+\lambda_{\phi\eta_1} v^2\right),\\
\mu_{\eta_2}^2&=& \frac{\lambda_{\eta_1\eta_2}^{\prime} u_1^3+\lambda_{\eta_1\eta_2} u_1^2 u_2 + 2 \lambda_{\eta_2} {u_2}^3+ \lambda_{\phi\eta_2}{u_2} v^2}{2 u_2}.
\end{eqnarray}
The CP even neutral scalar mass matrix, for neutral components of $\Phi, \eta_{1,2}$  is
\begin{equation}
\mathcal{M}^2_{\rm even}=\left(
\begin{array}{ccc}
 2 \lambda_{\phi} v^2 & \lambda_{\phi\eta_1} u_1 v & \lambda_{\phi\eta_2} u_2 v \\
 \lambda_{\phi\eta_1} u_1 v & 2 \lambda_{\eta_1} u_1^2+\frac{3 \lambda_{\eta_1\eta_2}^\prime u_1 u_2}{2} & \frac{3 \lambda_{\eta_1\eta_2}^\prime u_1^2}{2}+\lambda_{\eta_1\eta_2} u_1 u_2 \\
 \lambda_{\phi\eta_2} u_2 v & \frac{3 \lambda_{\eta_1\eta_2} u_1^2}{2}+\lambda_{\eta_1\eta_2} u_1 u_2 & 2 \lambda_{\eta_2} u_2^2-\frac{\lambda_{\eta_1\eta_2}^\prime u_1^3}{2 u_2} \\
\end{array}
\right).
\end{equation}
The CP odd neutral scalar mass matrix (for singlet scalar components) is
\begin{eqnarray}
\mathcal{M}^2_{\rm odd}=\left(
\begin{array}{cc}
 \frac{-9}{2} \lambda_{\eta_1\eta_2}^\prime u_1 u_2 & \frac{3}{2} \lambda_{\eta_1\eta_2}^\prime u^2_1 \\
 \frac{3}{2} \lambda_{\eta_1\eta_2}^\prime u_1^2 & -\frac{\lambda_{\eta_1\eta_2}^\prime u_1^3}{2 u_2} \\
\end{array}
\right)
\end{eqnarray}
with vanishing determinant leading to a Goldstone mode.
The charged scalar mass matrix, comprising of charged components of doublet $\zeta$ and singlet $\rho$, is
\begin{eqnarray}
\mathcal{M}^2_{\rm charged}=\left(
\begin{array}{cc}
 \mu_{\zeta}^2+\frac{1}{2} \left(\lambda_{\zeta\eta_1} u_1^2+\lambda_{\zeta\eta_2} u_2^2+\lambda_{\phi\zeta} v^2\right) & \frac{\lambda u_1 v}{2} \\
 \frac{\lambda u_1 v}{2} & \mu_{\rho}^2  + \frac{1}{2} \left(\lambda_{\eta_1\rho} u_1^2+\lambda_{\eta_2\rho} u_2^2+\lambda_{\phi\rho} v^2\right) \\
\end{array}
\right).
\end{eqnarray}
The neutral components of scalar doublet $\zeta$ acquire masses as
\begin{eqnarray}
M_{\zeta_R}^2=M_{\zeta_I}^2=\frac{1}{2} \left(2 \mu_{\zeta}^2 + \lambda_{\zeta\eta_1} u_1^2+\lambda_{\zeta\eta_2} u_2^2+\lambda_{\phi\zeta} v^2\right).
\end{eqnarray}
Other relevant couplings and mass relations are summarised below.
\begin{eqnarray}
\lambda_{\eta_1\eta_2}^\prime &=& -\frac{2 M_{A}^2 u_2}{u_1 \left(u_1^2+9 u_2^2\right)},\\
u_1&=&\frac{2 \cos\theta_{\rm ch} \left(M_{\zeta_R}^2-M_{\chi_1^+}^2 \right)}{\lambda \sin\theta_{\rm ch} \, v},\\
M_{\chi_2^+}&=&\sqrt{\frac{M_{\zeta_R}^2-\cos\theta_{\rm ch}^2 M_{\chi_1^+}^2}{\sin\theta_{\rm ch}^2}},\\ \nonumber
\mu_{\rho}^2&=& \frac{1}{{2 \lambda^2 \sin\theta_{\rm ch}^2 v^2}}\Big(-2 \cos\theta_{\rm ch}^4 M_{\chi_1^+}^2 \left(-4\lambda_{ \eta_1\rho} M_{\zeta_R}^2+4 \lambda_{\eta_1\rho} M_{\chi_1^+}^2 \sin\theta_{\rm ch}^2+\lambda^2 v^2\right)+\\ \nonumber
&& 2 \cos\theta_{\rm ch}^2 \left(\lambda^2 M_{\zeta_R}^2 v^2-2 \lambda_{\eta_1\rho} \left(M_{\zeta_R}^2-M_{\chi_1^+}^2 \sin\theta_{\rm ch}^2\right)^2\right)-4 \cos\theta_{\rm ch}^6 \lambda_{\eta_1\rho} M_{\chi_1^+}^4+\\ 
&& \lambda^2 \sin\theta_{\rm ch}^2 v^2 \left(2 M_{\chi_1^+}^2 \sin\theta_{\rm ch}^2-\lambda_{\eta_2\rho} u_2^2-\lambda_{\phi\rho} v^2\right)\Big),\\
\lambda_{\phi}&=&\frac{\cos\theta_{13}^2 \left(\cos\theta_{12}^2 M_{H_1}^2+ M_{H_2}^2\sin\theta_{12}^2\right)+M_{H_3}^2 \sin\theta_{13}^2}{2 v^2},\\ \nonumber
\lambda_{\eta_1} &=& \frac{1}{4 u_1^2}\Big(2 \sin\theta_{23}^2 \left(\sin\theta_{13}^2 \left(\cos\theta_{12}^2 M_{H_1}^2+M_{H_2}^2 \sin\theta_{12}^2\right)+\cos\theta_{13}^2 M_{H_3}^2\right)+\\ \nonumber
&&2 \cos\theta_{23}^2 \left(\cos\theta_{12}^2 M_{H_2}^2+M_{H_1}^2 \sin\theta_{12}^2\right)+\\
&&4 \cos\theta_{12} \cos\theta_{23} \sin\theta_{12} \sin\theta_{13} \sin\theta_{23} (M_{H_1}^2-M_{H_2}^2)-3 \lambda_{\eta_1\eta_2}^\prime u_1 u_2\Big),\\
\lambda_{\eta_2}&=&\frac{1}{4 u_2^3}\Big( \lambda_{\eta_1\eta_2}^\prime u_1^3 + 2 u_2 \Big(\cos\theta_{23}^2 \left(\sin\theta_{13}^2 \left(\cos\theta_{12}^2 M_{H_1}^2+ M_{H_2}^2 \sin\theta_{12}^2\right)+\cos\theta_{13}^2  M_{H_3}^2\right)+\\ \nonumber
&& \sin\theta_{23}^2 \left(\cos\theta_{12}^2  M_{H_2}^2+ M_{H_1}^2 \sin\theta_{12}^2\right)+2 \cos\theta_{12} \cos\theta_{23} \sin\theta_{12} \sin\theta_{13} \sin\theta_{23} \left( M_{H_2}^2- M_{H_1}^2\right)\Big)\Big),\\ \nonumber
\lambda_{\phi\eta_1} &=& \frac{1}{u_1 v}\Big(\cos\theta_{13} \sin\theta_{13} \sin\theta_{23} \left(-\cos\theta_{12}^2  M_{H_1}^2- M_{H_2}^2 \sin\theta_{12}^2+ M_{H_3}^2\right)+\\
&& \cos\theta_{12} \cos\theta_{13} \cos\theta_{23} \sin\theta_{12} \left( M_{H_2}^2- M_{H_1}^2\right)\Big),\\ \nonumber
\lambda_{\phi\eta_2} &=& \frac{1}{u_2 v}\Big(\cos\theta_{13} \big(\cos\theta_{23} \sin\theta_{13} \left(-\cos\theta_{12}^2  M_{H_1}^2- M_{H_2}^2 \sin\theta_{12}^2+ M_{H_3}^2\right)+\\
&& \cos\theta_{12} \sin\theta_{12} \sin\theta_{23} \left( M_{H_1}^2- M_{H_2}^2\right)\big)\Big),\\ \nonumber
\lambda_{\eta_1\eta_2} &=& -\frac{1}{2 u_1 u_2}\Big(2 \cos\theta_{12}^2 \cos\theta_{23} \sin\theta_{23} \left( M_{H_2}^2- M_{H_1}^2 \sin\theta_{13}^2\right)+2 \cos\theta_{12} \sin\theta_{12} \sin\theta_{13} (\cos\theta_{23}-\sin\theta_{23})\\ \nonumber
&&  (\cos\theta_{23}+\sin\theta_{23}) \left( M_{H_2}^2- M_{H_1}^2\right)-2 \cos\theta_{13}^2 \cos\theta_{23}  M_{H_3}^2 \sin\theta_{23}+2 \cos\theta_{23} \sin\theta_{12}^2 \sin\theta_{23}\\ 
&&  \left( M_{H_1}^2- M_{H_2}^2 \sin\theta_{13}^2\right)+3 \lambda_{\eta_1\eta_2}^\prime u_1^2\Big).
\end{eqnarray}
In the above expressions $\theta_{12}, \theta_{12}, \theta_{23}$ are mixing angles of neutral CP even scalar mass matrix with mass eigenvalues denoted by $M_{H_1}, M_{H_2}, M_{H_3}$ respectively.

\hspace*{\fill}

\providecommand{\href}[2]{#2}\begingroup\raggedright\endgroup

\end{document}